RU-96-12-B
%
%
\headline{\hfil \folio}
\hoffset=0.5truein
\hsize=5truein
\vsize=8truein
%
\catcode`@=11                           
\newskip\ttglue
\def\ninefonts{%
   \global\font\ninerm=cmr9%
   \global\font\ninei=cmmi9%
   \global\font\ninesy=cmsy9%
   \global\font\nineex=cmex10%
   \global\font\ninebf=cmbx9%
   \global\font\ninesl=cmsl9%
   \global\font\ninett=cmtt9%
   \global\font\nineit=cmti9%
   \skewchar\ninei='177%
   \skewchar\ninesy='60%
   \hyphenchar\ninett=-1%
   \moreninefonts
   \gdef\ninefonts{\relax}}%
\def\moreninefonts{\relax}                      


\def\elevenfonts{%
   \global\font\elevenrm=cmr10 scaled \magstephalf%
   \global\font\eleveni=cmmi10 scaled \magstephalf%
   \global\font\elevensy=cmsy10 scaled \magstephalf%
   \global\font\elevenex=cmex10%
   \global\font\elevenbf=cmbx10 scaled \magstephalf%
   \global\font\elevensl=cmsl10 scaled \magstephalf%
   \global\font\eleventt=cmtt10 scaled \magstephalf%
   \global\font\elevenit=cmti10 scaled \magstephalf%
   \global\font\elevenss=cmss10 scaled \magstephalf%
   \skewchar\eleveni='177%
   \skewchar\elevensy='60%
   \hyphenchar\eleventt=-1%
   \moreelevenfonts
   \gdef\elevenfonts{\relax}}%
\def\moreelevenfonts{\relax}

\def\twelvefonts{%
   \global\font\twelverm=cmr10 scaled \magstep1%
   \global\font\twelvei=cmmi10 scaled \magstep1%
   \global\font\twelvesy=cmsy10 scaled \magstep1%
   \global\font\twelveex=cmex10 scaled \magstep1%
   \global\font\twelvebf=cmbx10 scaled \magstep1%
   \global\font\twelvesl=cmsl10 scaled \magstep1%
   \global\font\twelvett=cmtt10 scaled \magstep1%
   \global\font\twelveit=cmti10 scaled \magstep1%
   \global\font\twelvess=cmss10 scaled \magstep1%
   \skewchar\twelvei='177%
   \skewchar\twelvesy='60%
   \hyphenchar\twelvett=-1%
   \moretwelvefonts
   \gdef\twelvefonts{\relax}}%
\def\moretwelvefonts{\relax}                    

\def\fourteenfonts{%
   \global\font\fourteenrm=cmr10 scaled \magstep2%
   \global\font\fourteeni=cmmi10 scaled \magstep2%
   \global\font\fourteensy=cmsy10 scaled \magstep2%
   \global\font\fourteenex=cmex10 scaled \magstep2%
   \global\font\fourteenbf=cmbx10 scaled \magstep2%
   \global\font\fourteensl=cmsl10 scaled \magstep2%
   \global\font\fourteenit=cmti10 scaled \magstep2%
   \global\font\fourteenss=cmss10 scaled \magstep2%
   \skewchar\fourteeni='177%
   \skewchar\fourteensy='60%
   \morefourteenfonts
   \gdef\fourteenfonts{\relax}}%
\def\morefourteenfonts{\relax}                  


\def\tenmibfonts{
   \global\font\tenmib=cmmib10%
   \global\font\tenbsy=cmbsy10%
   \skewchar\tenmib='177%
   \skewchar\tenbsy='60%
   \gdef\tenmibfonts{\relax}}

\def\elevenmibfonts{
   \global\font\elevenmib=cmmib10 scaled \magstephalf%
   \global\font\elevenbsy=cmbsy10 scaled \magstephalf%
   \skewchar\elevenmib='177%
   \skewchar\elevenbsy='60%
   \gdef\elevenmibfonts{\relax}}

\def\twelvemibfonts{
   \global\font\twelvemib=cmmib10 scaled \magstep1%
   \global\font\twelvebsy=cmbsy10 scaled \magstep1%
   \skewchar\twelvemib='177%
   \skewchar\twelvebsy='60%
   \gdef\twelvemibfonts{\relax}}

\def\fourteenmibfonts{
   \global\font\fourteenmib=cmmib10 scaled \magstep2%
   \global\font\fourteenbsy=cmbsy10 scaled \magstep2%
   \skewchar\fourteenmib='177%
   \skewchar\fourteenbsy='60%
   \gdef\fourteenmibfonts{\relax}}

\def\mib{
   \tenmibfonts%
   \textfont0=\tenbf\scriptfont0=\sevenbf%
   \scriptscriptfont0=\fivebf%
   \textfont1=\tenmib\scriptfont1=\seveni%
   \scriptscriptfont1=\fivei%
   \textfont2=\tenbsy\scriptfont2=\sevensy%
   \scriptscriptfont2=\fivesy}%

\def\ninepoint{\ninefonts               
   \def\rm{\fam0\ninerm}%
   \textfont0=\ninerm\scriptfont0=\sevenrm\scriptscriptfont0=\fiverm
   \textfont1=\ninei\scriptfont1=\seveni\scriptscriptfont1=\fivei
   \textfont2=\ninesy\scriptfont2=\sevensy\scriptscriptfont2=\fivesy
   \textfont3=\nineex\scriptfont3=\nineex\scriptscriptfont3=\nineex
   \textfont\itfam=\nineit\def\it{\fam\itfam\nineit}%
   \textfont\slfam=\ninesl\def\sl{\fam\slfam\ninesl}%
   \textfont\ttfam=\ninett\def\tt{\fam\ttfam\ninett}%
   \textfont\bffam=\ninebf
   \scriptfont\bffam=\sevenbf
   \scriptscriptfont\bffam=\fivebf\def\bf{\fam\bffam\ninebf}%
   \def\mib{\relax}%
   \tt\ttglue=.5emplus.25emminus.15em
   \normalbaselineskip=11pt
   \setbox\strutbox=\hbox{\vrule height 8pt depth 3pt width 0pt}%
   \normalbaselines\rm\singlespaced}%

\def\tenpoint{
   \def\rm{\fam0\tenrm}%
   \textfont0=\tenrm\scriptfont0=\sevenrm\scriptscriptfont0=\fiverm
   \textfont1=\teni\scriptfont1=\seveni\scriptscriptfont1=\fivei
   \textfont2=\tensy\scriptfont2=\sevensy\scriptscriptfont2=\fivesy
   \textfont3=\tenex\scriptfont3=\tenex\scriptscriptfont3=\tenex
   \textfont\itfam=\tenit\def\it{\fam\itfam\tenit}%
   \textfont\slfam=\tensl\def\sl{\fam\slfam\tensl}%
   \textfont\ttfam=\tentt\def\tt{\fam\ttfam\tentt}%
   \textfont\bffam=\tenbf
   \scriptfont\bffam=\sevenbf
   \scriptscriptfont\bffam=\fivebf\def\bf{\fam\bffam\tenbf}%
   \def\mib{%
      \tenmibfonts%
      \textfont0=\tenbf\scriptfont0=\sevenbf%
      \scriptscriptfont0=\fivebf%
      \textfont1=\tenmib\scriptfont1=\seveni%
      \scriptscriptfont1=\fivei%
      \textfont2=\tenbsy\scriptfont2=\sevensy%
      \scriptscriptfont2=\fivesy}%
   \tt\ttglue=.5emplus.25emminus.15em
   \normalbaselineskip=12pt
   \setbox\strutbox=\hbox{\vrule height 8.5pt depth 3.5pt width 0pt}%
   \normalbaselines\rm\singlespaced}%

\def\elevenpoint{\elevenfonts           
   \def\rm{\fam0\elevenrm}%
   \textfont0=\elevenrm\scriptfont0=\sevenrm\scriptscriptfont0=\fiverm
   \textfont1=\eleveni\scriptfont1=\seveni\scriptscriptfont1=\fivei
   \textfont2=\elevensy\scriptfont2=\sevensy\scriptscriptfont2=\fivesy
   \textfont3=\elevenex\scriptfont3=\elevenex\scriptscriptfont3=\elevenex
   \textfont\itfam=\elevenit\def\it{\fam\itfam\elevenit}%
   \textfont\slfam=\elevensl\def\sl{\fam\slfam\elevensl}%
   \textfont\ttfam=\eleventt\def\tt{\fam\ttfam\eleventt}%
   \textfont\bffam=\elevenbf
   \scriptfont\bffam=\sevenbf
   \scriptscriptfont\bffam=\fivebf\def\bf{\fam\bffam\elevenbf}%
   \def\mib{%
      \elevenmibfonts%
      \textfont0=\elevenbf\scriptfont0=\sevenbf%
      \scriptscriptfont0=\fivebf%
      \textfont1=\elevenmib\scriptfont1=\seveni%
      \scriptscriptfont1=\fivei%
      \textfont2=\elevenbsy\scriptfont2=\sevensy%
      \scriptscriptfont2=\fivesy}%
   \tt\ttglue=.5emplus.25emminus.15em
   \normalbaselineskip=13pt
   \setbox\strutbox=\hbox{\vrule height 9pt depth 4pt width 0pt}%
   \normalbaselines\rm\singlespaced}%

\def\twelvepoint{\twelvefonts\ninefonts 
   \def\rm{\fam0\twelverm}%
   \textfont0=\twelverm\scriptfont0=\ninerm\scriptscriptfont0=\sevenrm
   \textfont1=\twelvei\scriptfont1=\ninei\scriptscriptfont1=\seveni
   \textfont2=\twelvesy\scriptfont2=\ninesy\scriptscriptfont2=\sevensy
   \textfont3=\twelveex\scriptfont3=\twelveex\scriptscriptfont3=\twelveex
   \textfont\itfam=\twelveit\def\it{\fam\itfam\twelveit}%
   \textfont\slfam=\twelvesl\def\sl{\fam\slfam\twelvesl}%
   \textfont\ttfam=\twelvett\def\tt{\fam\ttfam\twelvett}%
   \textfont\bffam=\twelvebf
   \scriptfont\bffam=\ninebf
   \scriptscriptfont\bffam=\sevenbf\def\bf{\fam\bffam\twelvebf}%
   \def\mib{%
      \twelvemibfonts\tenmibfonts%
      \textfont0=\twelvebf\scriptfont0=\ninebf%
      \scriptscriptfont0=\sevenbf%
      \textfont1=\twelvemib\scriptfont1=\ninei%
      \scriptscriptfont1=\seveni%
      \textfont2=\twelvebsy\scriptfont2=\ninesy%
      \scriptscriptfont2=\sevensy}%
   \tt\ttglue=.5emplus.25emminus.15em
   \normalbaselineskip=14pt
   \setbox\strutbox=\hbox{\vrule height 10pt depth 4pt width 0pt}%
   \normalbaselines\rm\singlespaced}%

\def\fourteenpoint{\fourteenfonts\twelvefonts 
   \def\rm{\fam0\fourteenrm}%
   \textfont0=\fourteenrm\scriptfont0=\twelverm\scriptscriptfont0=\tenrm
   \textfont1=\fourteeni\scriptfont1=\twelvei\scriptscriptfont1=\teni
   \textfont2=\fourteensy\scriptfont2=\twelvesy\scriptscriptfont2=\tensy
   \textfont3=\fourteenex\scriptfont3=\fourteenex
      \scriptscriptfont3=\fourteenex
   \textfont\itfam=\fourteenit\def\it{\fam\itfam\fourteenit}%
   \textfont\slfam=\fourteensl\def\sl{\fam\slfam\fourteensl}%
   \textfont\bffam=\fourteenbf
   \scriptfont\bffam=\twelvebf
   \scriptscriptfont\bffam=\tenbf\def\bf{\fam\bffam\fourteenbf}%
   \def\mib{%
      \fourteenmibfonts\twelvemibfonts\tenmibfonts%
      \textfont0=\fourteenbf\scriptfont0=\twelvebf%
      \scriptscriptfont0=\tenbf%
      \textfont1=\fourteenmib\scriptfont1=\twelvemib%
      \scriptscriptfont1=\tenmib%
      \textfont2=\fourteenbsy\scriptfont2=\tenbsy%
      \scriptscriptfont2=\tenbsy}%
   \normalbaselineskip=17pt
   \setbox\strutbox=\hbox{\vrule height 12pt depth 5pt width 0pt}%
   \normalbaselines\rm\singlespaced}%
%
%

\def\singlespaced{
   \baselineskip=\normalbaselineskip}           


%
%
\twelvepoint
%
%
\def\begintitle{\begingroup%
\obeylines\fourteenpoint\bf\parindent=0.29truein}
\def\endtitle{\vglue 1truecm\endgroup}

\def\showheadline#1#2{\headline={\ifnum\pageno>1{\ifodd\pageno{\hfil\tenpoint #1\hfil} %
\else{\hfil\tenpoint #2\hfil}\fi} \else{\hfil}\fi}}

\def\address#1{\hbox to \hsize{\hglue 0.29in\relax
\vbox{\hsize=4.70in\relax\rightskip=0pt plus 1in\relax\noindent#1}\hfil}}

\long\def\beginaddress#1\endaddress{\vglue 6pt\address{#1}\vglue 24pt}

\def\beginabstract{\begingroup\leftskip=0.29in%
\tenpoint\noindent{\bf Abstract\ \ \ }}
\def\endabstract{\vskip 1pt minus1pt\endgroup}

\def\finalversion{\headline{\hfil}}

\def\section#1{\vskip 24pt plus4pt minus4pt\goodbreak\leftline{\bf #1}%
\vglue 12pt\nobreak\noindent\kern -0.0em}

\def\subsection#1{\vskip 12pt plus4pt minus4pt\goodbreak\leftline{\bf #1}%
\nobreak\noindent\kern -0.0em}

\def\subsubsection#1{\vskip 12pt plus4pt minus4pt\goodbreak\leftline{\it #1}%
\nobreak\noindent\kern -0.0em}

\def\begincaption#1{\begingroup\tenpoint\noindent#1\ \ \ }
\def\endcaption{\endgroup}

\newbox\@capbox                                 
\newcount\@caplines                             

\def\references{\section{REFERENCES}\tenpoint\parindent=0pt
\raggedright\rightskip=0pt plus 5em}

\def\ref#1#2{\hbox to \hsize{\vbox{\tenpoint\hsize=0.2in\relax #1\hfil}
\hfil\vtop{\hsize=4.75in\relax\tenpoint #2}}}
%
%
\vglue 1.0truein

\finalversion
%
%
\input epsf
%
\overfullrule 0pt
%
%
%
%
%
\begintitle
\centerline {\bf{A POSSIBLE MECHANISM FOR}} 
\centerline {\bf{PSEUDO-GAP IN CUPRATES}}
\endtitle
\centerline {Hai-cang Ren}
\centerline {Department of Physics, The Rockefeller University}
\centerline {New York, NY 10021-6399, U.S.A.}
\bigskip
            
\beginabstract
The electron spectrum in the normal phase of cuprates is analyzed in terms 
of the boson-fermion model. It is argued that the existence of the uncondensed 
pairs and the quasi-two-dimensional nature of the crystal structure 
are responsible for the pseudo-gap above $T_C$ observed recently. 
\endabstract

\section{1. Introduction}

Recently, ARPES experiments revealed a novel property of the 
underdoped cuprate superconductors in their normal phase. There is 
a pseudo-gap in the electron spectrum, i.e. a suppression of the 
spectral weight near the Fermi level above the transition temperature 
[1]. It is speculated by a number of authors that the phenomena may be 
attributed to the existence of the uncondensed pairs in the normal 
phase. The present article is an attempt in this direction.

A common feature of the cuprate superconductors is that the coherence 
length is comparable with the lattice spacing on the $CuO_2$ plane. 
Therefore the Cooper pairs, if they exist, are highly localized in the 
coordinate space and consequently can be regarded as boson degrees of 
freedom. The superconductivity is thereby associated with the 
kinematical Bose-Einstein condensation of these pairing states and 
there must be uncondensed pairs in the normal phase. 

The crystal structure of all cuprate materials is highly anisotropic
It consists of parallel layers of $CuO_2$, the minimum distance between 
oxygens (where the doped charge carriers may reside) on the same layer 
is less than $3\AA$. The separations between neighboring $CuO_2$ layers 
are not even but periodic with an average spacing around $6\AA$. The number 
of layers in each period varies from one for $La_{2-x}Sr_xCuO_4$ 
($T_C\leq40{\rm K}$) to three for $Tl_2Ca_2Ba_2Cu_3O_x$ ($T_C\leq120{\rm K}$).
Although the coherence perpendicular to the $CuO_2$ 
layer is required for the condensate to form below $T_C$, the normal phase 
properties of the under-doped materials are largely two dimensional in 
nature. This has been confirmed by a number of experiments on the transport 
coefficients of cuprates [2]. 

The Bose-Einstein condensation is characterized by the relation that 
the thermal wavelength of the bosons, $$\lambda_T=\sqrt{{2\pi \over m_b
\kappa T}}\eqno(1.1)$$ with $m_b$ the effective mass of bosons, at the 
transition temperature $T=T_C$  should be comparable 
with the inter-boson distance, $l$, so that quantum mechanical coherence 
takes place. The ratio ${\lambda_T\over l}$ is 1.38 for an ideal Bose 
gas and 1.65 for $HeII$ at the $\lambda$-transition. For high $T_C$ materials, 
the ratio $${\lambda_{T_C}\over l}\sim 2.8\eqno(1.2)$$
was estimated in [3] based on the results of the $\mu SR$ experiment [4]. 
The reasoning went as follows: According to the experimental results, 
the transition temperature $T_C$ of the under-doped material is inversely 
proportional to the square of the magnetic penetration depth with a constant 
of proportionality universal for all cuprates. If the supercurrent is 
identified with the superfluid of bosons. A relation 
$$T_C({\rm K})=10^5{m_e\over m_b}{n_b\over d}\eqno(1.3)$$ 
was obtained, where $m_e$ is the electron mass in vacuum, $m_b$ is the
in-plane effective mass of the bosons, $n_b$ is the density of the bosons 
(counted as number of bosons per $\AA^2$ and per layer) and $d$ the 
average separation between $CuO_2$ layers (in the unit of $\AA$). 
Combining the relation (1.3) with $d=6\AA$, $n_b=l^{-2}$ and the definition 
of the thermal wavelength (1.1), the ratio (1.2) follows independent of 
$m_b$. This ratio is certainly consistent with the picture of Bose-Einstein 
condensation.

In an isotropic situation, a relation $\lambda_{T_C}\sim l$ implies $T_C\propto
n_b^{2\over 3}$ in accordance with (1.1) and $n_b=l^{-3}$. This is not the 
case of (1.3). Therefore we have to resort to the quasi-two dimensional 
layer structure of the system.
Consider an ideal Bose gas in quasi-two dimensions whose kinetic energy 
consists a free motion term in $x-y$ plane and a hopping(tunneling) 
term in $z$-direction, i.e.
$$\omega_{\vec p,K}={p^2\over 2m_b}+2t_b(1-\cos Kd),\eqno(1.4)$$
where $\vec p$ is the momentum in $x-y$ plane, $K$ is the Bloch momentum 
in $z$-direction and $t_b$ is the out-of-plane 
hopping amplitude. Expanding the cosine of (1.4) to $K^2$ term, we obtain 
the out-of-plane effective mass $$M_b=(2t_bd^2)^{-1}\eqno(1.5)$$
The density of states of the spectrum (1.4) is
$$\rho(\omega)={m_b\over 2\pi}\cases{1 & for $\omega>4t_b$;\cr {2\over\pi}
\sin^{-1}\sqrt{{\omega\over 4t_b}} & for $\omega<4t_b$.\cr}.\eqno(1.6)$$
The boson density (counted as number of bosons per unit area and per layer) 
at a temperature $T$ and a chemical potential $\mu<0$ 
is $$n_b=\int_{-\pi}^\pi{d\theta\over 2\pi}\int {d^2p\over (2\pi)^2}
{1\over e^{\beta(\omega-\mu)}-1}={1\over\lambda_T^2}
\int_{-\pi}^\pi {d\theta\over 2\pi}
\ln{1\over 1-e^{-\beta[2t_b(1-\cos\theta)-\mu]}},\eqno(1.7)$$ 
where $\beta={1\over\kappa T}$, and $\lambda_T=
\sqrt{{2\pi\over m_b\kappa T}}$ is the thermal wavelength
of the bosons in the $x-y$ plane. For $\beta t_b<<1$ and $-\beta\mu<<1$, 
the integral (1.7) may be expressed in terms of elementary functions. We 
find that $$n_b={1\over\lambda_T^2}\ln{2\kappa T\over (2t_b-\mu)+
\sqrt{(2t_b-\mu)^2-4t_b^2}}.\eqno(1.8)$$ At the transition $\mu=0$ and we 
have $$n_b={1\over\lambda_{T_C}^2}\ln{\kappa T_C\over t_b}
={m_b\kappa T_C\over 2\pi}\ln{\kappa T_C\over t_b}.\eqno(1.9)$$
Regarding $x-y$ plane as a $CuO_2$ layer, it was found in [5] that this 
quasi-linear relation between $n_b$ and 
$T_C$ fits the $\mu SR$ result (1.3) remarkably well with 
$$\ln {\kappa T_C\over t_b}\sim 8 \hbox{ at $T_C=100K$ },\eqno(1.10)$$
which implies that $$t_b/\kappa T_C=3.6\times 10^{-4}.\eqno(1.11)$$
Furthermore, the formula (1.10) can be generalized to the material with 
multi-layers in a period in $z$ direction provide we replace the hopping 
amplitude $t_b$ in (1.10) by the geometrical mean of the hopping amplitudes 
between each pair of neighboring layers within one period. This also 
explains the universality of the $n_b-T_C$ profile (1.3), since the 
tunneling nature of boson motion in $z$-direction makes the geometrical 
mean of the hopping amplitudes sensitive only to the average separation 
between the $CuO_2$ layers and this average is approximately the same 
for different cuprates.
	
The large logarithmic factor is obviously caused by the large fluctuations 
of low-lying bosons, i.e. the 2$D$ nature of the density of state (1.6) 
down to the tiny energy scale $t_b$. Since the square root of the 
logarithm of (1.10) gives the quoted ratio (1.2), the quasi-two-dimensional 
character also responsible for the factor two enlargement of this ratio 
compared with that of an ideal Bose gas in three dimensions, bearing in 
mind that within 3$D$, the change of this ratio from an ideal Bose gas 
to the highly interaction $HeII$ is merely $20\%$. 

A simple phenomenological model dealing with system with localized Cooper 
pairs was proposed in [3] and [6], which will be refered to as the 
boson-fermion model. In the rest of the paper, we shall 
first review this model and then apply it to explore the consequences 
of the presence of uncondensed bosons within a quasi-two-dimensional crystal 
in the normal phase.

\section{2. The Boson-Fermion Model in Quasi-Two Dimensions}
 
The boson-fermion model appropriate to the quasi-two dimensional crystal 
structure is formulated in ref.[7]. The grand 
Hamiltonian of the system reads 
$$H=\sum_{\vec P,s}\epsilon_{\vec P}a_{\vec Ps}^\dagger a_{\vec Ps}
+\sum_{\vec P}\omega_{\vec P}b_{\vec P}^\dagger b_{\vec P}$$
$$+{g\over \sqrt{{\cal N}\Omega}}\sum_{\vec P,\vec Q}
(b_{\vec P}a_{{\vec P\over 2}+\vec Q\uparrow}^\dagger
a_{{\vec P\over 2}-\vec Q\downarrow}^\dagger+
b_{\vec P}^\dagger a_{{\vec P\over 2}-\vec Q\downarrow}
a_{{\vec P\over 2}-\vec Q\uparrow}),\eqno(2.1)$$ where
$a_{\vec Ps}$ and $a_{\vec Ps}^\dagger$ ($b_{\vec P}$ and 
$b_{\vec P}^\dagger$) are annihilation and creation operators of 
electrons (bosons), the subscript $s$ denotes the spin orientation 
($s=\uparrow$ or $\downarrow$), 
$$\epsilon_{\vec P}={p^2\over 2m_f}+2t_f(1-\cos Kd)-\mu,\eqno(2.2)$$
$$\omega_{\vec P}={p^2\over 2m_b}+2t_b(1-\cos Kd)+2(\nu-\mu),\eqno(2.3)$$
$m_f(m_b)$ is the in-plane effective mass of electrons(bosons), $t_f(t_b)$
is the out-of-plane hopping amplitude of fermions(bosons), $\mu$ is the 
chemical potential of the system, $2\nu$ is the energy of a static boson 
relative to two static electrons ($2\nu=\omega_{\vec P}|_{\vec P=0}-
2\epsilon_{\vec P}|_{\vec P=0}$), $\Omega$ is the total area of a $CuO_2$ 
layer and ${\cal N}$ is the total number of layers of the whole system. 
The vectors with capital symbols,
$\vec P$ and $\vec Q$ are all three dimensional, i.e. 
$$\vec P=\vec p+K\hat z$$ and $$\vec Q=\vec q+K^\prime\hat z$$ 
with $\vec p$, $\vec q$ parallel to the $x-y$ plane, $K$, 
$K^\prime$ Bloch momenta in $z$ direction whose value varies between 
$-\pi/d$ and $\pi/d$. The conserved electric charge is given by
$$N=\sum_{\vec P,s}a_{\vec Ps}^\dagger a_{\vec Ps}
+2\sum_{\vec P}b_{\vec P}^\dagger b_{\vec P}.\eqno(2.4)$$
For resonant bosons, $\nu>0$ and the dimensionless coupling constant $g$ 
is related to the boson width in vacuum through
$$\Gamma={g^2\over 4}m_f\cases{1 & for $\nu>4t_f$;\cr {2\over\pi}\sin^{-1}
\sqrt{{\nu\over 4t_f}}& otherwise.\cr}$$ For $g=0$, the model 
implies an ideal Bose gas and an ideal fermion gas in chemical equilibrium. 
In what follows we shall assume that $\nu>4t_f$. The effective expansion 
parameter of the perturbation series is $\hat g$ defined by   
$$\hat g^2={\Gamma\over \nu}=g^2{m_f\over 4\nu}.\eqno(2.5)$$

The condensation mechanism with resonance bosons can be understood 
easily in the weak coupling limit, $\hat g<<1$ and at $T=0$. Starting 
with low charge density, the fermionic levels $\epsilon_{\vec P}$ are 
filled first. The bosons are unstable, existing in virtual state through 
the interaction and the condensate is of BCS type.  With the increasing 
charge density, the fermion levels are filled up to $\nu$. Additional 
charges will be stabilized at the zero momentum boson level to minimize the 
energy since the decay channels of a static bosons are all locked by 
the Fermi sea and the condensate is of Bose-Einstein type. 

In the super-phase, we introduce a long range order $B$ representing the 
Bose condensate at the $\vec P=0$, i.e.
$$b_{\vec P}|_{\vec P=0}=\sqrt{\Omega}Be^{i\alpha}+b^\prime,\eqno(2.6)$$ 
where $B>0$, $\alpha$ is a constant phase, and then $$H=H_0+H_1,\eqno(2.7)$$ 
$$H_0=2\Omega(\nu-\mu)+\sum_{\vec P}\omega_{\vec P}b_{\vec P}^\dagger 
b_{\vec P}$$ $$+\sum_{\vec P}[\epsilon_{\vec P}(a_{\vec P\uparrow}^\dagger 
a_{\vec P\uparrow}+a_{\vec P\downarrow}^\dagger a_{\vec P\downarrow})
+\Delta(e^{i\alpha}a_{\vec P\uparrow}^\dagger a_{-\vec P\downarrow}^\dagger+
e^{-i\alpha}a_{-\vec P\downarrow}a_{\vec P\uparrow})]\eqno(2.8)$$
with $\Delta=gB$, and $H_1$ includes the cubic terms in $a$'s and $b$'s. 
The electron spectrum to the zeroth order in $H_1$ but to 
all orders in $\Delta$ can be obtained via a Bogoliubov transformation 
of $H_0$. In order to compare notes with the normal phase results later on, 
we adapt diagrammatic technique to derive the electron spectrum. All the 
diagrams throughout this paper are thermal with discrete imaginary energies 
(Matsubara variables) and continuous spatial momenta. The real-time
(real energy) Green's functions can be obtained by replacing 
all external Matsubara energies by real and continuous energies. The 
uniqueness of such a replacement was established in [8].

Returning to $H_0$, we associate a solid line with the bare 
electron propagator $${i\over i\nu_n-\epsilon_{\vec P}}\eqno(2.9)$$
with $\nu_n={2\pi\over\beta}(n+{1\over 2})$ and $\epsilon_{\vec P}$ given 
by (2.2), a cross with two incoming fermion lines is associated with a 
factor $-i\Delta$ and that with two outgoing lines with a factor $i\Delta$.
The electron propagator to all orders in $\Delta$ but to zeroth order in 
$H_1$ (represented by a solid line with two arrows), with a momentum 
$\vec P$ and a discrete energy $i\nu_n$, $S_n(\vec P)$, 
corresponds to the sum of the diagrams in Fig.1. 

\topinsert
\hbox to\hsize{\hss
	\epsfxsize=4.0truein\epsffile{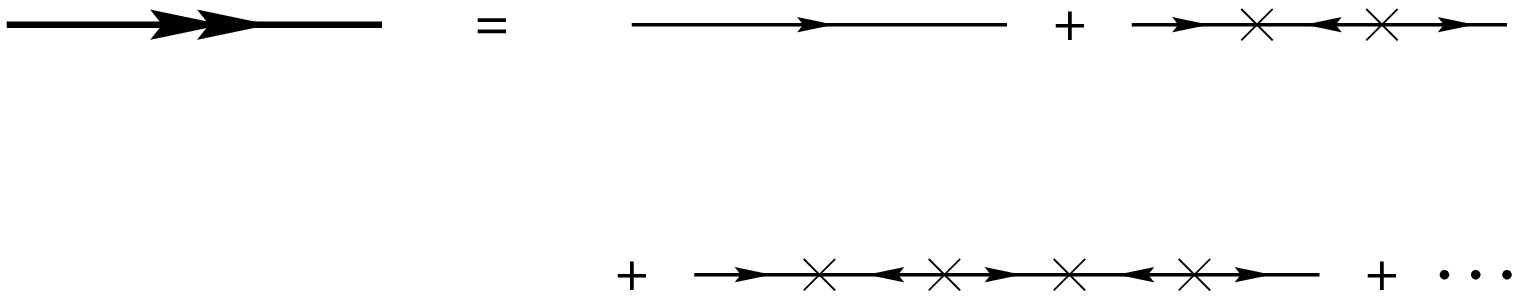}
	\hss}
\begincaption{Figure 1.}
The diagrammatic expansion of the electron propagator in the super phase
\endcaption
\endinsert

\noindent
We have $$S_n(\vec P)={i\over i\nu_n-\epsilon_{\vec P}}\sum_{N=0}^\infty
(-i\Delta)^N(i\Delta)^N\Big({i\over -i\nu_n-\epsilon_{\vec P}}\Big)^N
\Big({i\over i\nu_n-\epsilon_{\vec P}}\Big)^N$$ $$
={\nu_n-i\epsilon_{\vec P}\over \nu_n^2+E_{\vec P}^2}\eqno(2.10)$$
where $E_{\vec P}=\sqrt{\epsilon_{\vec P}^2+\Delta^2}$.
Replacing the Matsubara energy $i\nu_n$ by $p_0+i0^+$ with $p_0$ real, 
we obtain the retarded electron propagator 
$$S_R(p_0,\vec P)={i(p_0+\epsilon_{\vec P})\over (p_0+i0^+)^2-
E_{\vec P}^2}.\eqno(2.11)$$ The corresponding spectral density is 
$$A(p_0,\vec P)\equiv{1\over\pi}{\rm Re}S_R(p_0,\vec P)={|p_0+\epsilon_{\vec
P}|\over 2E_{\vec P}}[\delta(p_0-E_{\vec P})+\delta(p_0+E_{\vec P})]
\eqno(2.12)$$ with two infinitely sharp quasi-particle peaks. 
On writing $$S(p_0,\vec P)={i\over p_0+i0^+-\epsilon_{\vec P}-\Sigma(p_0+i0^+
,\vec P)}\eqno(2.13)$$ we extract the self-energy function
$$\Sigma(p_0,\vec P)={\Delta^2\over p_0+\epsilon_{\vec P}}\eqno(2.14)$$

\section{3. The Electron Self-Energy in the Normal Phase}

Above the transition temperature, $T>T_C$, the coherence between different 
layers is no longer crucial and we shall approximate the model Hamiltonian 
(2.1) by a purely two-dimensional one, i.e.
$$H=\sum_{\vec p,s}\epsilon_{\vec p}a_{\vec ps}^\dagger a_{\vec ps}
+\sum_{\vec p}\omega_{\vec p}b_{\vec p}^\dagger b_{\vec p}$$
$$+{g\over \sqrt{\Omega}}\sum_{\vec p,\vec q}
(b_{\vec p}a_{{\vec p\over 2}+\vec q\uparrow}^\dagger
a_{{\vec p\over 2}-\vec q\downarrow}^\dagger+
b_{\vec p}^\dagger a_{{\vec p\over 2}-\vec q\downarrow}
a_{{\vec p\over 2}+\vec q\uparrow}),\eqno(3.1)$$ where 
$$\epsilon_{\vec p}={p^2\over 2m_f}-\mu,\eqno(3.2)$$
$$\omega_{\vec p}={p^2\over 2m_b}+\delta\eqno(3.3)$$
with $\delta=2(\nu-\mu)$ whose value in the vicinity above $T_C$ is 
regarded of the same order of $t_b$ in (1.11). All 
momenta are two-dimensional from now on. 

\bigskip
\noindent 
{\it{3.1 The self-energy to one-loop order}}
\indent

In terms of the thermal diagrams, a free electron propagator, represented 
by a solid line, with a momentum $\vec p$ and an energy 
$i{2\pi\over\beta}(n+{1\over 2})$ corresponds 
to $${i\over i{2\pi\over\beta}(n+{1\over 2})-\epsilon_{\vec p}}\eqno(3.4)$$ 
with $n$ an integers; 
a free boson propagator, represented by a dashed line, with a momentum 
$\vec p$ and an energy $i{2\pi\over\beta}n$
corresponds to $${i\over i{2\pi\over\beta}n-\omega_{\vec p}};\eqno(3.5)$$  
a vertex with two incoming electrons and one outgoing boson corresponds to 
$-ig$ and that with two outgoing electrons and one incoming boson to $ig$. 
The energy and momentum are conserved at each vertex. A loop corresponds
to a sum and an integral $${i\over\beta}\sum_l\int {d^2q\over (2\pi)^2}$$
with $l$ the integer labeling the Matsubara energy of a chosen line of the loop 
and $\vec q$ the momentum of that line. A minus sign is associated with 
a fermionic loop. All arrows in the diagrams follow the charge flow.

The electron self-energy to the one-loop order is given by $i$ times the 
amputated part of the diagram in Fig. 2a, i.e.

\topinsert
\hbox to\hsize{\hss
	\epsfxsize=4.0truein\epsffile{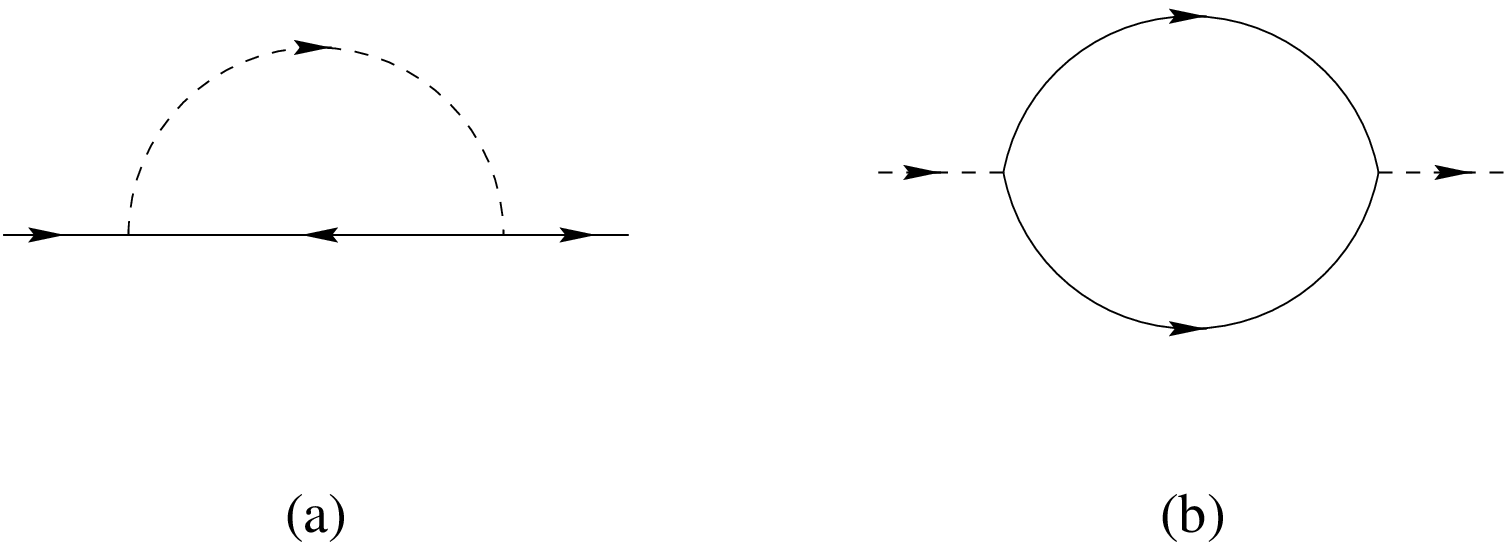}
	\hss}
\begincaption{Figure 2.}
The one-loop diagrams for the self-energy of an electron (a) and the 
self-energy of a boson (b).
\endcaption
\endinsert

$$\Sigma_n(\vec p)=i{i\over\beta}(-ig)ig\sum_l\int{d^2q\over (2\pi)^2}
\Big[{i\over i{2\pi\over\beta}l-\omega_{\vec q}}\Big]
\Big[{i\over i{2\pi\over\beta}(l-n-{1\over 2})-\omega_{\vec q-\vec p}}\Big]
$$ $$=g^2\int{d^2q\over (2\pi)^2}{N_b(\vec q)+N_f(\vec q-\vec p)
\over i{2\pi\over\beta}(n+{1\over 2})-\omega_{\vec q}
-\epsilon_{\vec q-\vec p}},\eqno(3.6)$$
where $N_b$ and $N_f$ are boson and fermion distribution functions given by 
$$N_b(\vec p)={1\over e^{\beta\omega_{\vec p}}-1}\eqno(3.7)$$ and
$$N_f(\vec p)={1\over e^{\beta\epsilon_{\vec p}}+1}\eqno(3.8)$$ 
respectively. 
With the Matsubara energy $i{2\pi\over\beta}(n+{1\over 2})$ replaced by 
$p_0+i0^+$, we obtain the retarded self-energy function
$$\Sigma(p_0,\vec p)=g^2\int{d^2q\over (2\pi)^2}{N_b(\vec q)+N_f(\vec q-\vec p)
\over p_0-\omega_{\vec q}-\epsilon_{\vec q-\vec p}+i0^+}.\eqno(3.9)$$
If the bosonic chemical potential $-\delta$ were zero, the integration over
$\vec q$ would be logarithmically divergent. Therefore, in the limit 
$\delta\to 0$, we expect
$$\Sigma(p_0,\vec p)\to {\bar\Delta^2\over p_0+\epsilon_{\vec p}}
\eqno(3.10)$$ with
$$\bar\Delta^2={g^2m_b\kappa T\over 2\pi}\ln{\kappa T\over\delta}.
\eqno(3.11)$$ The right hand side of (3.10) is of the same form as (2.14) 
and a perfect gap emerges. With a moderate logarithmic factor, the gap 
becomes a pseudo one as smeared by the imaginary part of 
$\Sigma(p_0,\vec p)$. The integration in (3.9) can be carried out 
analytically in the situation $\mu>>\kappa T$ and $\mu>>\epsilon_{\vec p}$ 
and $\mu>>p_0$. The result reads
$$\Sigma(p_0,\vec p)=g^2{m_b\over 2\pi}\Bigg\{{\kappa T\over 
\sqrt{4r\mu\delta+(p_0+\epsilon_{\vec p}-\delta)^2}}\Bigg[
{\rm sign}(p_0+\epsilon_{\vec p}-\delta)$$ $$\times\ln{\sqrt{4r\mu\delta
+(p_0+\epsilon_{\vec p}-\delta)^2}+|p_0+\epsilon_{\vec p}-\delta|
\over \sqrt{4r\mu\delta+(p_0+\epsilon_{\vec p}-\delta)^2}-
|p_0+\epsilon_{\vec p}-\delta|}-i\pi\Bigg]$$ $$-{\ln r\over r-1} 
+{1\over 2}\sqrt{{\pi\kappa T\over 
r\mu}}\Big[f(e^{-\beta(p_0-\delta)})-if(e^{\beta(p_0-\delta)})
-i\zeta\Big({1\over 2}\Big)\Big]\Bigg\},\eqno(3.12)$$
where the function $f$ is defined as
$$f(z)={2\over\sqrt{\pi}}\int_0^\infty dx\sqrt{x}{ze^{-x}\over (1+ze^{-x})^2}
,\eqno(3.13)$$ $r=m_b/m_f$ and $\zeta(1/2)=-1.4604$. 
For $\vec p$ on the Fermi surface, $p=p_F$, $\epsilon_{\vec p}=0$ and we 
find that $\Sigma(p_0,\vec p)|_{p=p_F}=u(p_0)+iv(p_0)$ where
$$u(p_0)=g^2{m_b\over 2\pi}\Bigg[{\kappa T{\rm sign}(p_0-\delta)\over
\sqrt{4r\mu\delta+(p_0-\delta)^2}}\ln{\sqrt{4r\mu\delta
+(p_0-\delta)^2}+|p_0-\delta|\over \sqrt{4r\mu\delta+(p_0-\delta)^2}-
|p_0-\delta|}$$ $$-{\ln r\over r-1}
+{1\over 2}f(e^{-\beta(p_0-\delta)})\sqrt{{\pi\kappa T\over 
r\mu}}\Bigg]\eqno(3.14)$$ and $$v(p_0)=-g^2{m_b\over 2\pi}\Bigg[
{\pi\kappa T\over\sqrt{4r\mu\delta+(p_0-\delta)^2}}$$ with 
$$+{1\over 2}\Big(f(e^{\beta(p_0-\delta)})+\zeta\Big({1\over 2}\Big)\Big)
\sqrt{{\pi\kappa T\over r\mu}}\Bigg].\eqno(3.15)$$
The spectral function at the Fermi surface is
$$A(p_0,\vec p)|_{p=p_F}={1\over\pi}
{\rm Re}{i\over p_0-u(p_0)-iv(p_0)+u(\delta)},\eqno(3.16)$$
where we have renormalized the chemical potential in 
(3.16) by subtracting the constant term $u(\delta)$. The function 
$A(p_0,\vec p)$ on the Fermi surface is plotted in Fig.3 for $\ln{\kappa
T\over\delta}=8$, 6 and 4 respectively. The opening of a pseudo 
gap with increasing $\ln\kappa T/\delta$ is clearly shown. 

\topinsert
\hbox to\hsize{\hss
	\epsfxsize=4.0truein\epsffile{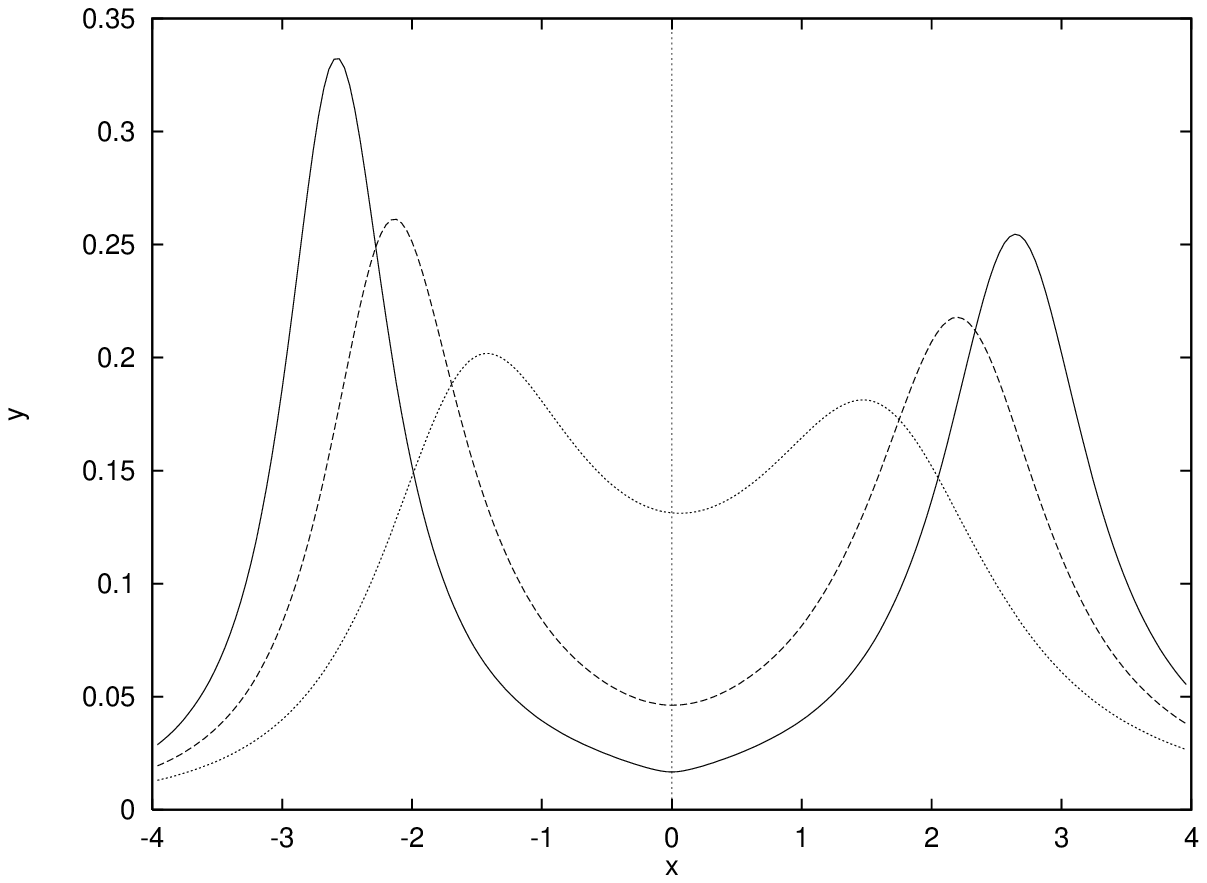}
	\hss}
\begincaption{Figure 3}
The spectral function at $\epsilon_{\vec p}=0$, with $\beta\mu=10$, 
$r=2$ and $g^2m_b\beta/2\pi=1$. The solid line corresponds to 
$\ln\kappa T/\delta=8$, the dashed line to $\ln\kappa T/\delta=6$ and 
the dotted line to $\ln\kappa T/\delta=4$, where $x=p_0/\kappa T$ and 
$y=A(p_0,\vec p)|_{p=p_F}\kappa T$.
\endcaption
\endinsert

\bigskip
\noindent 
{\it{3.2 A resummation of higher order diagrams}}
\indent

The validity of the perturbative results for $\Sigma(p_0,\vec p)$ 
requires both $\hat g^2<<1$ and $\hat g^2\ln{\kappa T\over \delta}<<1$. 
Nonperturbative effects will enter the game if the second inequality 
fails. In what follows, we shall examine the higher order diagrams and 
derive an expression which can be extended to the region $\hat g^2<<1$ 
but $\hat g^2\ln{\kappa T\over \delta}\sim 1$

A diagrams of an electron propagator to an arbitrary order consists of at 
least a string of electron lines, refered to as the main string, joining 
the initial and the final states. A number of vertices has to be attached 
to this line. In what follows, a vertex with an outgoing boson line will 
be refered to as a source and that with an incoming boson line will be 
refered to as a sink. On account of the charge conservation, there should be 
an even number of vertices with sources and sinks alternatively attached to 
the main string. Without decorating the boson lines, there are no electron 
lines outside the main string. Unlike the super phase (Fig. 1), in which 
the boson lines all terminated at the condensate, a boson line emitted from 
a source can land at any sink of the diagram. Therefore there are $N$! such 
diagrams to the $2N$-th order in $g$ and $N$ loops in each diagram. Such an 
expansion to 6th order in $g$ is displayed in Fig. 4. Each loop contains a
boson

\topinsert
\hbox to\hsize{\hss
	\epsfxsize=4.0truein\epsffile{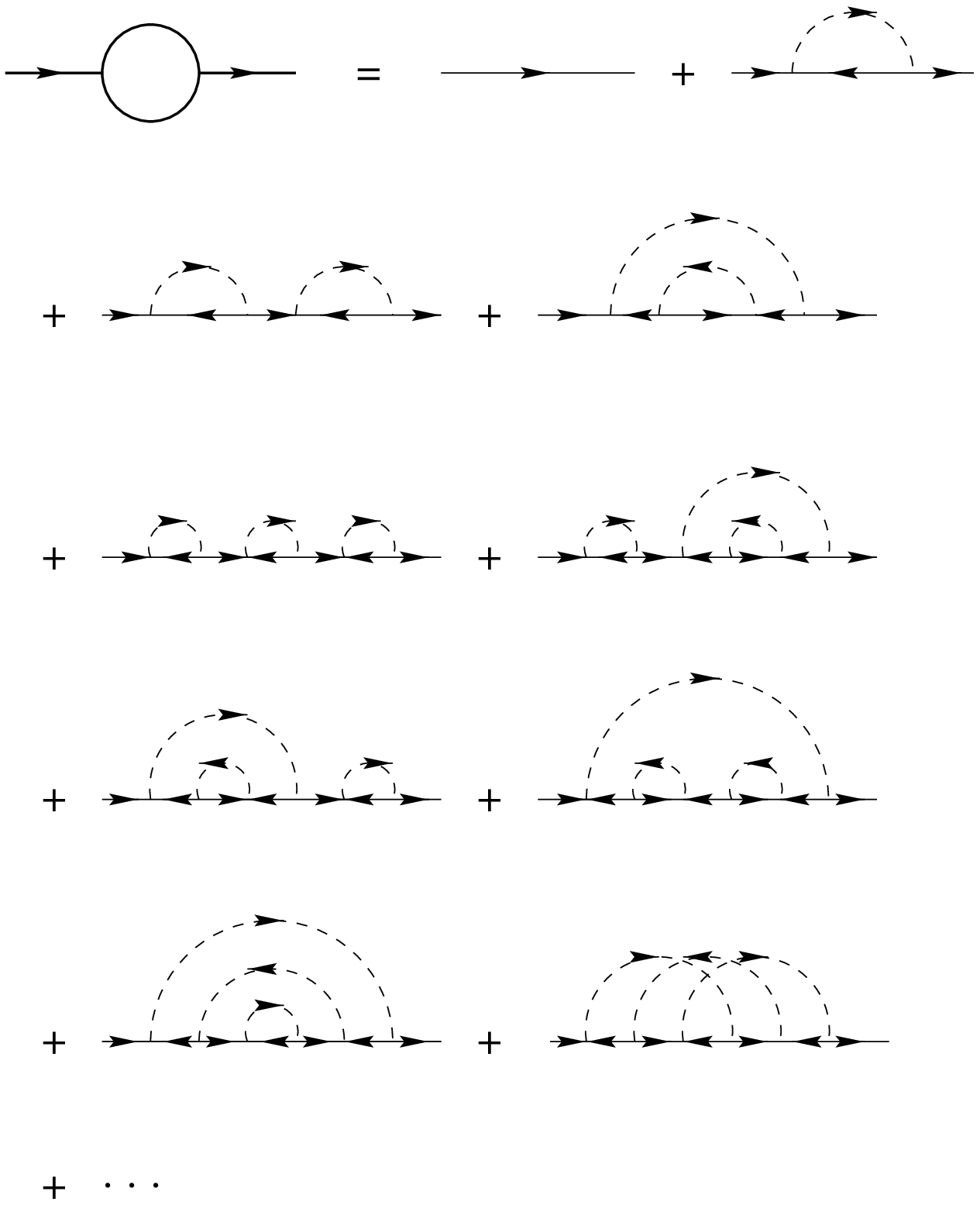}
	\hss}
\begincaption{Figure 4.}
The diagrammatic expansion of the electron propagator in the normal phase.
\endcaption
\endinsert

\noindent
leaving the main string and then coming back. The Matsubara energies 
of these bosons are denoted by $i{2\pi\over\beta}l_m$ ($m=1,2,...,N$)
with $l_m$ integers and the corresponding momenta by $\vec q_m$.
$l_m$'s are to be summed and $\vec q_m$'s are to be integrated. The leading 
divergence as $\delta\to 0$ comes from the integrals 
over $\vec q_m$'s of the term with $l_1=l_2=...=l_N=0$, where each momentum 
integral over $\vec q_m$ is logarithmically divergent. For a tiny $\delta$, 
each integral remains dominated by the lower limit, and contribute the 
amplitude of the diagram a factor $\ln{\Lambda\over \delta}>>1$ with $\Lambda$
a cutoff energy $\sim \kappa T$. To the leading order of these logarithms, 
the terms with some of $l_m$'s nonzero can be dropped and the deviation of 
the momentum of an internal electron line along the main string from the 
external momentum $\vec p$ due to the boson emissions and absorptions can be 
neglected. Collecting all the fermionic propagators and vertices with such 
an approximation, we find that all $2N$-th order diagrams 
contribute equally to the full electron propagator, $S_n(\vec p)$, a term 
$${i\over i\nu_n-\epsilon_{\vec p}}(-)^N\Big({\bar\Delta^2\over
\nu_n^2+\epsilon_{\vec p}^2}\Big)^N,\eqno(3.17)$$ where $\nu_n=
{2\pi\over\beta}(n+{1\over 2})$ and $$\bar\Delta^2=g^2{m_b\kappa T\over 2\pi}
\ln{\Lambda \over\delta}=2\hat g^2{m_b\over m_f}{\nu\kappa T\over \pi}
\ln{\Lambda \over\delta}$$ with $\hat g$ defined by (2.5). On summing over 
$N$, we obtain $$S_n(\vec p)={i\over i\nu_n-\epsilon_{\vec p}}
F\Big({\bar\Delta^2\over \nu_n^2+\epsilon_{\vec p}^2}\Big),\eqno(3.18)$$ 
where function $F$ is given by the asymptotic series
$$F(\zeta)\sim\sum_N(-)^NN!\zeta^N,\eqno(3.19)$$ which is formally 
Borel summable [9](It is advantageous in this case not to distinguish the 
one particle irreducible diagrams with the reducible ones). On substituting 
$$N!=\int_0^\infty dtt^Ne^{-t}\eqno(3.20)$$ into (3.19) and interchanging 
the order of the integration and the summation, we find
$$F(\zeta)=\int_0^\infty dt{e^{-t}\over 1+\zeta t}=-{1\over\zeta}
e^{1\over\zeta}{\rm Ei}\Big(-{1\over\zeta}\Big),\eqno(3.21)$$ 
where the function Ei is the exponential integral defined in [10]
$${\rm Ei}(z)\equiv\int_{-\infty}^zdt{e^t\over t}\eqno(3.22)$$ 
with ${\rm arg}(-z)<\pi$. ${\rm Ei}(z)$ is analytic on the $z$-plane cut 
along the positive real axis with the discontinuity across the cut given by
$${\rm Ei}(x+i0^+)-{\rm Ei}(x-i0^+)=-2\pi i.\eqno(3.23)$$ The asymptotic 
expansion with a large $|z|$ or a small $|z|$ are given respectively by
$${\rm Ei}(z)=\cases{\gamma+\ln(-z)+\sum_{n=1}^\infty{z^n\over n!n}&
for $|z|<<1$; \cr {e^z\over z}\sum_{n=0}^\infty{n!\over z^n} & 
for $|z|>>1$ \cr}\eqno(3.24)$$ with $\gamma$ the Euler constant. 

Replacing the imaginary energy $i\nu_n$ by $p_0+i0^+$, we obtain the 
retarded electron propagator:
$$S_R(p_0,\vec p)=i{p_0+\epsilon_{\vec p}\over \bar\Delta^2}
e^{-{p_0^2-\epsilon_{\vec p}^2\over\bar\Delta^2}}{\rm Ei}
\Big({(p_0+i0^+)^2-\epsilon_{\vec p}^2\over\bar\Delta^2}\Big),
\eqno(3.25)$$ which returns to the free propagator in the limit 
$\bar\Delta\to 0$, i.e.
$$\lim_{\bar\Delta\to 0}S_R(p_0,\vec p)={i\over p_0-\epsilon_{\vec p}+i0^+}.
\eqno(3.26)$$ The self-energy function extracted from (3.22) according to 
(2.13) reads $$\Sigma(p_0,\vec p)={\bar\Delta^2\over 
p_0+\epsilon_{\vec p}}\Big[z-z{d\over dz}\ln{\rm Ei}(z)\Big]\eqno(3.27)$$ 
with $z={p_0^2-\epsilon_{\vec p}^2\over \bar\Delta^2}$, which agrees with the 
one-loop result if $\hat g^2<<\hat g^2\ln{\kappa T\over \delta}<<1$. 

For a fixed $\vec p$, $S_R(p_0,\vec p)$ is an analytic 
function on the $p_0$ plane with two cuts running from $|\epsilon_{\vec p}|$ 
to $\infty$ and from $-|\epsilon_{\vec p}|$ to $-\infty$. The spectral 
function is given by the discontinuity across the cuts. It follows from 
(3.23) and (3.25) that $$A(p_0,\vec p)={1\over \pi}
{\rm Re}S_R(p_0+i0^+,\vec p)$$
$$=\cases{0,& for $|p_0|<|\epsilon_{\vec p}|$; \cr {|p_0+\epsilon_{\vec p}|
\over\bar\Delta^2}e^{-{p_0^2-\epsilon_{\vec p}^2\over\bar\Delta^2}},
& otherwise.\cr}\eqno(3.28)$$
The dependence of $A(p_0,\vec p)$ of (3.28) on the coupling constant $g$ 
in nonperturbative. On the Fermi surface, $\epsilon_{\vec p}=0$, we have
$$A(p_0,\vec p)|_{p=p_F}={|p_0|\over\bar\Delta^2}e^{-{p_0^2\over\bar\Delta^2}}
.\eqno(3.29)$$ Two peaks at $p_0=\pm\bar\Delta
/\sqrt{2}$ together with a depletion of states at the Fermi level 
$p_0=0$ corresponding to a pseudo-gap emerge.

\section{4. Comments}

In this final section, we shall comments on the validity of our results 
for a realistic system and its relation with known results in the 
literature.

The electron self-energy function in the boson-fermion model has been 
investigated in [11] and a pseudo-gap was found in $D=1$ and 2 by 
numerical solutions of the self-consistent equations for $\Sigma(p_0,\vec p)$. 
The approach taken in this article is purely analytical with emphasis on 
the quasi-2$D$ nature of the crystal structure. The one-loop 
result shows the trend of a pseudo-gap when the logarithmic factor 
$\ln{\kappa\over\delta}$ becomes sizable and the resummation of the 
diagrams in Fig. 4 gives the nonperturbative effect when the logarithmic 
factor becomes dominant. We notice that the entangled diagrams, e.g. the 
last one of Fig. 4, which is often neglected in the self-consistent 
method, contribute equally as the others to the leading order of the 
logarithm. Nevertheless, the result as to the existence of a pseudo-gap of 
the present work agrees with that of [11] qualitatively. 

The boson-fermion model is a phenomenological one, in which bosons are 
regarded as elementary. In a realistic system, bosons correspond to a 
pair of electrons and do not represent new degrees of freedom. This is not 
not a serious problem as long as the charge density is not too high to 
cause substantial overlapping among the pairs. A theorem proved in [12] 
states that for any fermionic Hamiltonian, there exist an equivalent 
boson-fermion Hamiltonian with the elementary bosons of the latter 
correspond to the fermion pairs of the former. For a Hubbard like model 
with an on-site attraction and a nearest neighbor repulsion, the low 
density behavior is indeed described by the boson-fermion model.

Although the motion of electrons and pairs in the $x-y$ plane is assumed 
to be in continuum, the result will not be changed if we replace the 
motion in $x-y$ plane by lattice hopping. Furthermore, the model 
Hamiltonian (2.1) can also accommodate a momentum dependent coupling $g$, 
say $$g\propto \hat p_x^2-\hat p_y^2,\eqno(4.1)$$
which results a pseudo-gap with $d$-wave character.

In the perturbative expansion (3.17), all boson propagators are kept 
undecorated. The decorations come in two ways: One is to 
replace the bare boson propagators by the dressed ones and the other 
is to include the scattering diagrams of these bosons. Both require
fermion loops in addition to the main string of fermion lines under 
discussion. 

The dressing of the boson propagator is through the self-energy 
function $\Pi_n(\vec p)$ defined at momentum $\vec p$ and Matsubara
energy $i{2\pi n\over\beta}$. The one loop diagram for $\Pi_n(\vec p)$ 
is depicted in Fig. 2b. We find, to that order and $n=0$ that
$$\Pi_n(\vec p)=a+bp^2\eqno(4.2)$$ as $\vec p\to 0$ with $a$ and $b$ 
functions of $T$ and $\mu$. The result is believed to be true in multi-loops. 
Since the contribution to (3.14) comes only from the boson propagator 
with zero Matsubara energy and low momentum, the effect of the boson self-
energy is merely a renormalization. The results of the last section is not 
modified if we regard $m_b$ and $\delta$ being renormalized quantities. 

The diagrams with boson scattering look problematic and the one of such 
diagrams to the lowest order in $g$ is shown in Fig. 5a. 

\topinsert
\hbox to\hsize{\hss
	\epsfxsize=4.0truein\epsffile{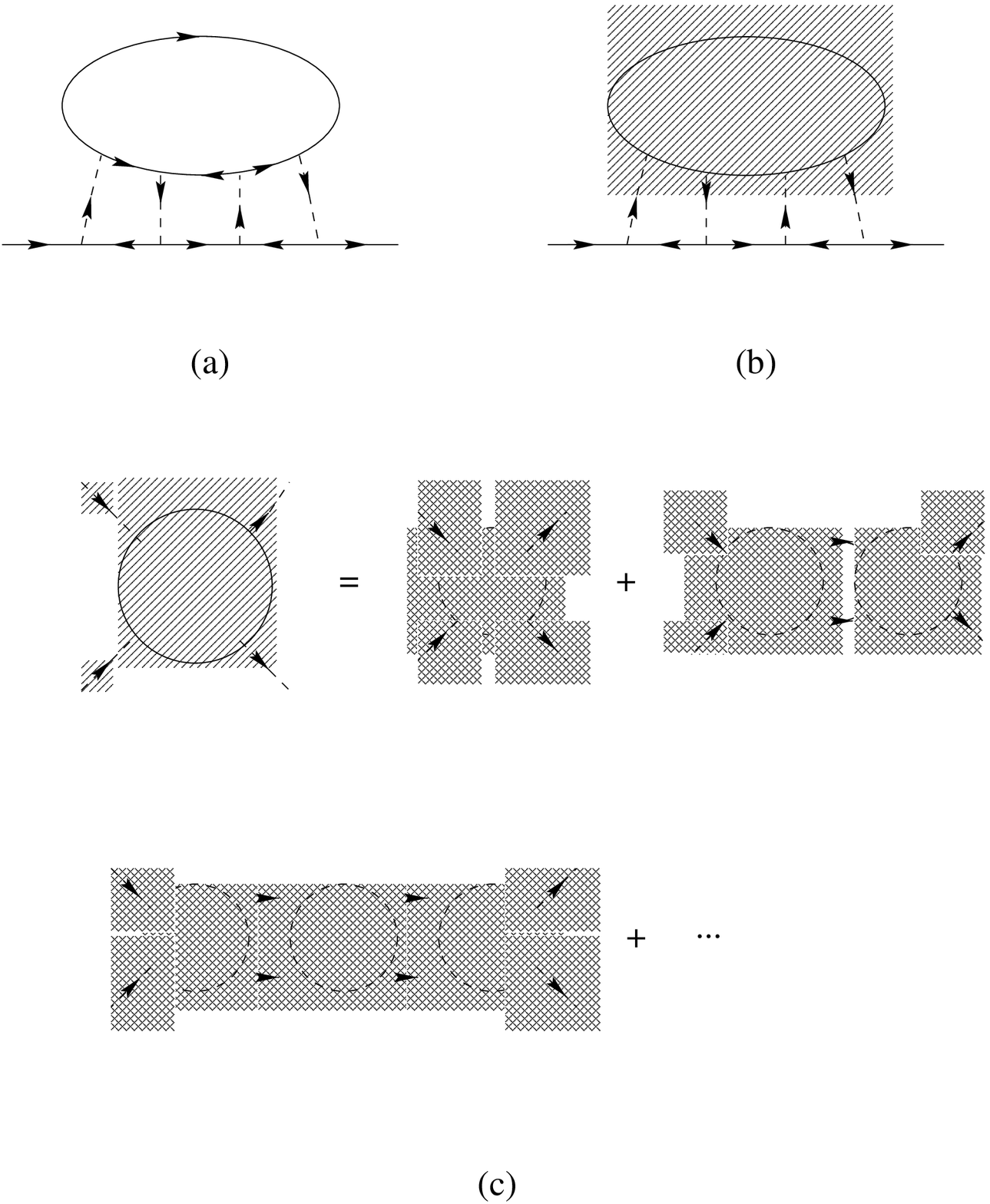}
	\hss}
\begincaption{Figure 5.}
(a) The lowest order diagram of the electron propagator with scattering of 
virtual bosons; (b) The propagator diagram with the scattering between the 
two virtual bosons to all orders; (c) The expansion of the shaded part of 
(b). A cross-shaded bubble is a two particle irreducible part.
\endcaption
\endinsert

\noindent 
A naive power counting indicates a linear divergence in the limit 
$\delta\to 0$. This however can be eliminated on summing over the diagrams 
with repeated scatterings between two bosons as is shown in Fig. 5b-c. 
Our result will survive if this is also true for multi-boson scatterings.

The Coulomb interaction has been left out throughout the discussions. 
Strictly speaking, the Coulomb interaction is perturbative only if the 
inter-particle distance is short compared with the Bohr radiu [13]. Therefore,
the results of the preceding section would be unaffected in a hypothetical 
situation when the charge density is high enough and the bosons are truly 
elementary. The situation with the 
real materials is, however, not in the perturbative region at all. On the 
other hand, the shielding of the Fermi sea truncates the long range tail of 
the direct Coulomb force and there are possibilities that our results 
survive. Diagrammatically, inclusion of the Coulomb 
interaction amounts to decorate the electron and fermion 
propagators and the boson-fermion vertices and to add scattering diagrams 
between electron and bosons. As long as the Fermi-liquid behavior of 
electron spectrum and the low momentum behavior of the boson spectrum
are not modified by Coulomb interactions alone, our result remain 
qualitatively intact.

Technically, the most drastic step leading to (3.25) is the Borel summation 
of the perturbative series (3.18). Is the sum unique?
An asymptotic expansion of a given function $F(\zeta)$, i.e. 
$$F(\zeta)\sim\sum_na_n\zeta^n,\eqno(4.3)$$
may well be the asymptotic expansion of another function, say 
$F(\zeta)+e^{-{1\over \zeta}}$. Therefore the sum of an asymptotic series is 
in general nonunique unless certain conditions are preimposed on the sum. 
According to the Watson theorem in [9], the sufficient conditions for 
the Borel sum to be a unique function $F(\zeta)$ which produce the asymptotic 
expansion of the right hand side of (4.3) are: 1) $F(\zeta)$ has to be 
analytic within the region $D$ defined by $|\zeta|<r$ and $|{\rm arg}(\zeta)|
<{\pi\over 2}+\alpha$ with $r>0$ and $\alpha>0$; 
2) $|a_n|=O(n!\sigma^n)$; with $\sigma>0$; 3)
$$|F(\zeta)-\sum_{n=0}^Na_n\zeta^n|=O((n+1)!\sigma^{n+1}r^{n+1}).\eqno(4.4)$$ 
uniformly within $D$. It follows from the rigorous spectral representation 
of the fermion propagator that the condition 1) is indeed satisfied if we 
identify $z$ with $\bar\Delta^2/ (\nu_n+\epsilon_{\vec p})$ and the 
coefficients of the perturbation 
series (3.18) meet 2). The condition 3) is difficult to justify 
nonperturbatively. Furthermore, each term in (3.18) is only a part of 
the perturbation in terms of $g$, since only the leading logarithmic term 
is reserved. Therefore the rigorous mathematical justification of the Borel 
sum (3.18) remains open.  

\section{Acknowledgement}

The author is indebted to Mr. O. Tchernyshyov for interesting him with 
this problem and for stimulating conversations. He is grateful to Professor 
N. N. Khuri for his critical reading of the manuscript. He would also like to 
thank Professor V. P. Nair for discussions and to thank Dr. James Liu for 
advice. This work is supported in part by the U. S. Department of Energy 
under Contract Grant DE-FG02-91ER40651.

\references
\ref{1.}{A. G. Loeser, D. S. Dessau and Z. X. Shen, {\it Physica} 
{\bf C263}, 208 (1996); H. Ding, {\it et. al}, {\it Nature}, 
{\bf Vol. 382}, 51 (1996).}
\ref{2.}{N. P. Ong, et. al., "Charge Transport Properties of Cuprate 
Superconductors", in {\it "High-$T_C$ Superconductivity and $C_{60}$ Family}, 
Proceeding of CCAST Symposium/Workshop, ed. S. Q. Feng and H. C. Ren, 
Gordon and Breach Pub. Inc., 1994; S. Uchda, "Optical Spectra of High-$T_C$ 
Superconductors", {\it ibid}.}
\ref{3.}{R. Friedberg and T. D. Lee, {\it Phys. Lett.} {\bf A138}, 423 
(1989); {\it Phys. Rev. B} {\bf 40}, 6745 (1989).}
\ref{4.}{Y. Uemura, et. al., {\it Phys. Rev. Lett. } {\bf 62}, 2317 (1989);
Y. Uemura, "Energy Scales of High $T_C$ Cuprates, Doped-Fullerenes, and 
Other Exotic Superconductors", in {\it "High-$T_C$ Superconductivity and 
$C_{60}$ Family}, Proceeding of CCAST Symposium/Workshop, ed. S. Q. Feng 
and H. C. Ren, Gordon and Breach Pub. Inc., 1994.}
\ref{5.}{R. Friedberg, T. D. Lee and H. C. Ren, {\it Phys. Lett. A} 
{\bf 152}, 423 (1991)}
\ref{6.}{R. Micnas, J. Ranninger and S. Robaszkiewicz, {\it Rev. Mod. Phys.} 
{\bf 62}, 113 (1990), and the references therein.}
\ref{7.}{R. Friedberg, T. D. Lee and H. C. Ren, {\it Phys. Lett. A} 
{\bf 152}, 417 (1991)}
\ref{8.}{G. Baym and N. D. Mermin, {\it J. Math. Phys.}, {\bf 2}, 232(1961)}
\ref{9.}{G. Hardy, {\it Divergent Series}, Oxford University Press, 
New York (1949).} 
\ref{10.}{N. N. Lebedev, {\it Special Functions and Their Applications}, 
Trans. R. Silverman, Dover Pub. Inc., 1972}
\ref{11.}{J. Ranninger, J. M. Robin and M. Eschrig, {\it Phys. Rev. Lett.}, 
{\bf 74}, 4027 (1995).}
\ref{12.}{R. Friedberg, T. D. Lee and H. C. Ren, {\it Phys. Rev. B} {\bf 50}, 
10190 (1994).}
\ref{13.}{M. Gell-Mann and K. A. Brueckner , {\it Phys. Rev. } {\bf 106}, 364,
(1957).}

\vfill\eject
\end

\bye